\documentclass[12pt,preprint]{aastex}

\newcommand{\gtsim}{\mbox
{{\raisebox{-0.4ex}{$\stackrel{>}{{\scriptstyle\sim}}$}}}}

\slugcomment{Revised version \today}

\shorttitle{Changes in jet speed and cone angle in SS433}
\shortauthors{Blundell \& Bowler}

\begin{document}

\title{Jet velocity in SS433: its anti-correlation with
  precession-cone angle and dependence on orbital phase}

\author{Katherine M.\ Blundell\altaffilmark{1} and Michael G.\
  Bowler\altaffilmark{1}} 

\altaffiltext{1}{University of Oxford, Department of Physics, Keble
  Road, Oxford, OX1 3RH, U.K.}

\begin{abstract}
We present a re-analysis of the optical spectroscopic data
on SS\,433 from the last quarter-century and demonstrate
that these data alone contain systematic and identifiable
deviations from the traditional kinematic model for the
jets: variations in speed, which agree with our analysis
of recent radio data; in precession-cone angle and in
phase. We present a simple technique for separating out the
jet speed from the angular properties of the jet axis,
assuming only that the jets are symmetric. With this
technique, the archival optical data reveal
that the variations in jet speed and in precession-cone
angle are anti-correlated in the sense that when faster jet
bolides are ejected the cone opening angle is smaller.  We
also find speed oscillations as a function of {\em orbital}
phase.
\end{abstract}


\section{Introduction}
\label{sec:intro}

In a recent paper \citep{Blu04} we presented the deepest yet radio
image of SS\,433, which revealed an historical record over two
complete precession periods of the geometry of the jets.
Detailed analysis of this image revealed systematic deviations from 
the standard kinematic model \citep{Mar84,Eik01}.  Variations
in jet speed, lasting for as long as tens of days, were needed to
match the detailed structure of each jet.  Remarkably, these
variations in speed were equal, matching the two jets
simultaneously.

\begin{figure}
\epsscale{0.5}
\plotone{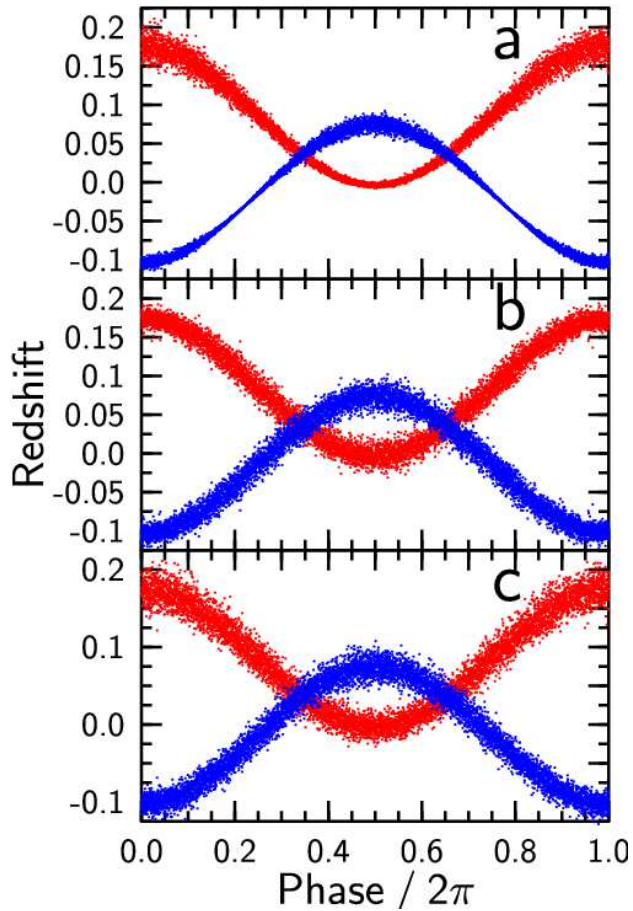}
\figcaption{\label{fig:varywithphase} Demonstration that the
  dispersion in redshift with phase is unaffected by velocity
  variations as long as they are accompanied by pointing variations:
   {\bf (a)} random variations in velocity drawn from
  a gaussian of half-width 0.013$c$, {\bf (b)} random variations in
  $\theta$ and in phase drawn from gaussian distributions of 2.7\,deg
  and 2.95 days respectively and {\bf (c)} uncorrelated variations in
  all three of the quantities above.
 } 
\end{figure}

The Doppler residuals to the kinematic model
show little variation with the precessional phase of the jets and this
observation rules out variations in jet speed {\em alone} as the
source of the residuals \citep[e.g.\, ][]{Kat82a,Eik01}.  Very little
phase variation is obtained if the pointing angle jitters
\citep{Kat82a} but there is no evidence excluding symmetric speed
variations of the magnitude reported in \cite{Blu04} superposed on
pointing jitter, Fig\,\ref{fig:varywithphase}.
Thus our findings from the radio image led us to re-analyse the
archival optical data.  G.\ Collins II and S.\ Eikenberry (with kind
permission of B.\ Margon) made available their compiled
datasets, published in \cite{Col00} and used in \cite{Eik01}.  We use
the Collins' compilation (available at
http://www-astro.physics.ox.ac.uk/$\sim$kmb/ss433/) because of its
higher quoted precision, but very similar results are obtained from
Margon's. 

\section{Speed and angular variations from the optical data}
\label{sec:variations}

If the precessing jet axis of SS\,433 traces out a cone of semi-angle
$\theta$ about a line which is oriented at an angle $i$ to our
line-of-sight with jet velocity $\beta$ in units of $c$ ($\gamma = (1
- \beta^2)^{-1/2}$), the redshifts measured from the west jet
($z_{+}$) and the east jet ($z_{-}$) are given, if the jets are
symmetric, by:
\begin{equation}
\label{eq:redshift}
  z_{\pm} = -1 + \gamma[1 \pm \beta\sin{\theta}\sin{i}\cos{\phi} \pm
  \beta\cos{\theta}\cos{i}],
\end{equation}
where $\phi$ is the phase of the precession cycle (see
http://www-astro.physics.ox.ac.uk/$\sim$kmb/ss433/).  Addition of
$z_{+}$ and $z_{-}$ in Eqn\,\ref{eq:redshift} gives an expression
relating the observed redshifts to the jet speed independently of any
angular variation.  Re-arrangement gives
\begin{equation}
\label{eq:sum}
  \beta = \left[ 1 - \left[1 + \frac{z_{+} +  z_{-}}{2}\right]^{-2}  \right] ^{1/2}.
\end{equation}
The quantity $z_{+} + z_{-}$ fluctuates very
substantially (see Fig\,\ref{fig:resids_v_time}). 
Subtraction of the expressions for $z_{+}$ and $z_{-}$ gives the angular properties $a$ of the orientation of the
jet axis, with the speed divided out using
Eqn\,\ref{eq:sum}:
\begin{equation}
\label{eq:ang}
a =   \frac{z_{+} -  z_{-}}{2\beta\gamma} = \sin{\theta}\sin{i}\cos{\phi}
  + \cos{\theta}\cos{i}, 
\end{equation}
  
Fluctuations in $\beta$, $\theta$ and $\phi$ are predominantly
symmetric, as described by Eqns 1--3, when fluctuations in $z_{+} +
z_{-}$ represent symmetric fluctuations in speed.  (The velocity
variations in the two radio jets \citep{Blu04} are highly
symmetric; the standard deviation on the difference in the speeds is
less than $0.004\,c$ and on the common velocity $0.014\,c$.)

The disadvantages of the variables $s = z_{+} + z_{-}$ and $a$
(Eqn\,\ref{eq:ang}) are that their interpretation is simple only for
perfect symmetry and that they may only be used for the 395 out of 486
observations which record a simultaneous pair.  Their merits are
exemplified by Fourier analyses of the time distributions.  We used
the algorithm of \cite{Rob87} which accounts for the uneven
time-sampling of the data.  The angular data $a$ clearly revealed
periodicities corresponding to the nodding of the precession axis
\citep{Kat82,New82,Col02} and the 162-day precession period, clearly
seen in Fig\,\ref{fig:ft}a.  There is no periodicity in the speed data
$s$ (Fig\,\ref{fig:ft}b) common to the angular data,
consistent with perfect symmetry.  The speed data also indicate a
periodicity at 13.08\,days (the periodicity at 12.58\,days matches a
beat with Earth's orbital period 365\,days); to investigate this, we
folded the data over 13.08\,days in 20 phase bins, and in each bin the
mean speed ($\beta$, from Eqn\,\ref{eq:sum}) was
derived. Fig\,\ref{fig:fold} shows a clear sinusoidal oscillation with
orbital phase.  The rms variation in speed which oscillates with
orbital phase is smaller by a factor of three than the overall speed
dispersion.  This oscillation with amplitude $2000\,{\rm km\,s^{-1}}$
may be because the speed with which the bolides are ejected is a
function of orbital phase, but the excursions in Fig\,\ref{fig:fold}
could also be interpreted as due to orbital motion; in that case
SS\,433's orbital speed is $\sim 400\, {\rm km\, s}^{-1}$.

\section{Anti-correlated deviations in jet speed and $\theta$}
\label{sec:correlations}

We fitted Collins' dataset with the kinematic model, including
nodding.  From our fit, we derived model redshift pairs and hence the
variables $s$ and $a$.  Subtraction of these model variables from
those constructed from the data gave residuals $\Delta s$ in $s$ and
$\Delta a$ in $a$.  The $\beta$ variation is shown in
Fig\,\ref{fig:resids_v_time}c.  The standard deviation of this
histogram is 0.013, in excellent agreement with the result from our
radio image (0.014).  Examples of the residuals in $s$ and in the
angular variable $a$ are plotted in Fig\,\ref{fig:resids_v_time}; the
variations in speed and angular residuals are anti-correlated.  From
Eqns\,\ref{eq:redshift}--\ref{eq:ang}, maintaining the assumption of
symmetry:

\begin{equation}
\label{eq:s_2}
   \Delta s^2  = 4 \beta^2 \gamma^6  \Delta
  \beta^2 ,
\end{equation}
\begin{eqnarray}
\label{eq:a_2}
   \Delta a^2  &=& (\cos\theta \sin{i} \cos\phi -
  \sin\theta \cos{i})^2   \Delta  \theta^2  +
  (\sin\theta \sin{i} \sin\phi)^2  \Delta\phi^2   \\ \nonumber
  &-&  2 \sin\theta \sin{i} \sin\phi (\cos\theta \sin{i} \cos\phi -
  \sin\theta \cos{i})   \Delta  \theta \Delta \phi ,
\end{eqnarray}
\begin{eqnarray}
\label{eq:as}
   \Delta a\,\Delta s  &=& 2 \beta \gamma^3 
[(\cos\theta \sin{i} \cos\phi - \sin\theta \cos{i}) 
  \Delta\beta\,\Delta\theta  \\ \nonumber
 &-& \sin\theta \sin{i} \sin\phi  \Delta\beta \Delta\phi],
\end{eqnarray}
where $\Delta \beta$, $\Delta \theta$ and $\Delta \phi$ represent the
variations in $\beta$, $\theta$ and $\phi$ respectively. Averaging
over many cycles for any given value of the 
phase $\phi$ yields the averages $\langle \Delta a^2 \rangle$,
$\langle \Delta s^2 \rangle$, $\langle \Delta a \Delta s \rangle$ as
functions of $\phi$, in terms of the parameters $\langle \Delta
\beta^2 \rangle$, $\langle \Delta \theta^2 \rangle$, $\langle
\Delta\beta\,\Delta\theta \rangle$ and so on.  The fit to $\langle
\Delta\beta\,\Delta\theta \rangle$ is shown in Fig\,\ref{fig:a_s}.
The parameters are given in Table\,\ref{tab:fits}; the global $\chi^2
/ NDF$ is 34.8/24, where $NDF$ is the number of degrees of freedom.  

Fig\,\ref{fig:a_s} shows that the quantity $\langle \Delta a\,\Delta s
\rangle$ has an almost pure cosinusoidal variation with phase, as
given by the first term on the right hand side of Eqn\,\ref{eq:as}.
This shape is the unique signature of a correlation between variations
in $\beta$ and $\theta$.  $\langle \Delta \beta^2 \rangle$ shows no
correlation with $\phi$ and $\langle \Delta a^2 \rangle$ very little.
The latter requires fluctuations in both $\theta$ and in $\phi$. We
remark that $\langle \Delta (z_{+} - z_{-})\,\Delta s \rangle $ does
not show a strong correlation with $\phi$; nor should it, using the
parameters from Table 1.  Removing the varying speed from $z_{+} -
z_{-}$ was crucial in revealing this correlation in $\langle \Delta
a\,\Delta s \rangle$.

\section{The redshift residual plot}
\label{sec:eikfigfive}

Consider the plane of redshift residuals, as in figure~5 of
\cite{Eik01}, with symmetric excursions from the kinematic model in
$\beta$, $\theta$ and $\phi$.  Comparison of their figure~5 (similar
to our Fig\,\ref{fig:eikfigfive}f) with our
Figs\,\ref{fig:eikfigfive}a and \ref{fig:eikfigfive}b requires the
presence of both angular variations (to spread the points along the
line $y = -x$) and velocity variations (to spread the points
perpendicular to this line --- note that even their quoted redshift
measurement error of 0.003, likely an over-estimate, will not account
for this breadth).  Inclusion of all these variations, correlated as
in Table\,\ref{tab:fits}, gives Fig\,\ref{fig:eikfigfive}d which
resembles that from the data (Fig\,\ref{fig:eikfigfive}f).  The
simulations in Fig\,\ref{fig:eikfigfive}d take no account of the
(stochastic) duration of the excursions; the duration of variations in
Fig\,\ref{fig:eikfigfive}e were drawn from a gaussian with half-width
2\,days.

Thus Figs\,\ref{fig:varywithphase} and \ref{fig:eikfigfive} establish
the consistency of the optical data with perfect symmetry and with
speed fluctuations whose magnitude is in excellent agreement with
those found in the radio image.  In addition, the slope of
Fig\,\ref{fig:eikfigfive}e is $-0.765 \pm 0.034$, and that for
Fig\,\ref{fig:eikfigfive}f from the Collins data set is $-0.786$ and
not $-1$; this long standing curiosity is explained by the physics
from the \S\,\ref{sec:correlations} fit.

\section{Other assumptions}
\label{sec:other}

If the jet speed were constant, the residuals to $z_{+}$ and $z_{-}$
would, for the case of strictly antiparallel jets, be correlated as in
Fig\,\ref{fig:eikfigfive}a; to spread the distribution of points
perpendicular to this line it is necessary to allow some independence
in the pointing of the two jets.  The variation with precessional
phase of the quantities $\langle \Delta s^2\rangle$, $\langle \Delta
a^2\rangle$ and $\langle \Delta a \Delta s\rangle$ is then almost as
well described ($\chi^2 / NDF = 36.3/24$) as by our fit of
\S\,\ref{sec:correlations}.  Such a model would not be able to explain
the radio jet morphology \citep{Blu04} and this fit required angular
fluctuations breaking symmetry to have rms values $\sim 1/2$ of those
preserving symmetry.  Angular jitter alone cannot account for the
slope of $\Delta z_{-}$ versus $\Delta z_{+}$
(Fig\,\ref{fig:eikfigfive}) differing from $-1$, unless the jitter in
the East jet is systematically smaller than in the West jet.
Velocity variation breaks symmetry in the Doppler shifts and accounts
naturally for this observation.  A better fit than either
($\chi^2/NDF$ = 25.5/21) was achieved by allowing some symmetry
breaking angular fluctuations in addition to symmetric velocity
fluctuations: in this case the rms symmetry breaking fluctuations were
$\sim 1/4$ of those preserving symmetry.  In all cases the rms
fluctuations in $\theta$ were $\sim 1/3$ of the rms fluctuations in
$\phi$, as would be expected for pointing angle fluctuations described
by \cite{Kat82a} as ``isotropic''.

\section{Concluding remarks}

\begin{figure}
\epsscale{1.0}
\plotone{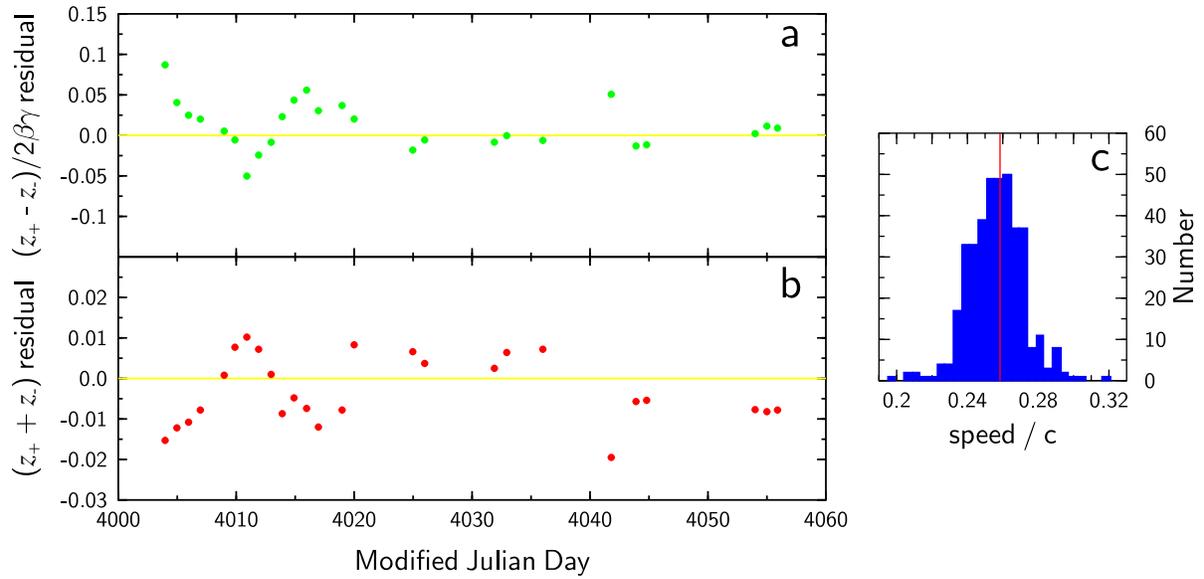}
\figcaption{\label{fig:resids_v_time} {\bf (a)} and {\bf (b)} examples
of angular and speed residuals versus time.  The variations suggest
anti-correlation, \S\,\ref{sec:correlations}.  {\bf (c)} Distribution
of speeds from the data compiled by Collins. The mean is indicated by
the vertical red line, and the standard deviation in $\beta$ is 0.013. }
\end{figure}

\begin{figure}
\epsscale{1.0}
\plotone{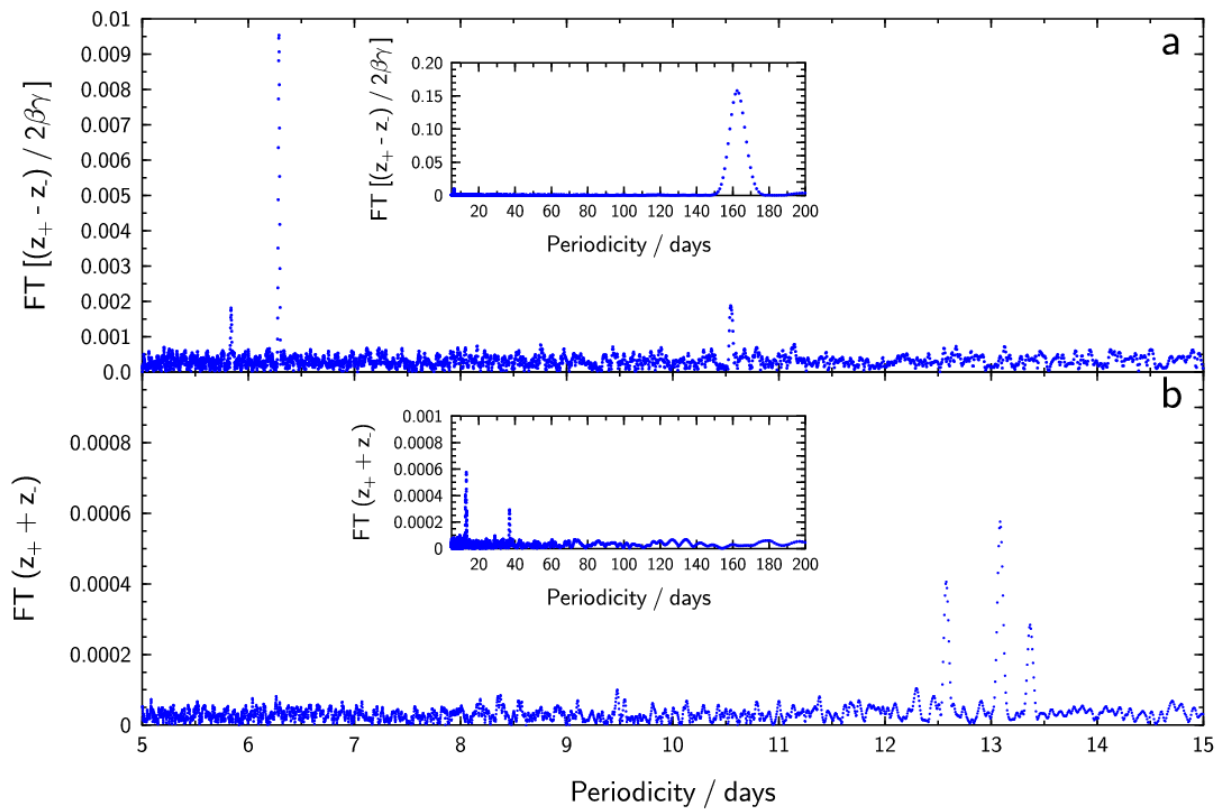}
\figcaption{\label{fig:ft} Fourier transform of data as described in
  \S\,\ref{sec:variations}.  {\bf (a)} The differences of all redshift
  pairs, divided by $2\beta \gamma$ (see Eqn\,\ref{eq:ang}) which, if
  the jets are symmetric, depend only on the angular properties of the
  jet.   {\bf (b)} The sums of all
  redshift pairs which, if the jets are symmetric, only depend on the
  speed of the jet.  }
\end{figure}

\begin{figure}
\epsscale{0.7}
\plotone{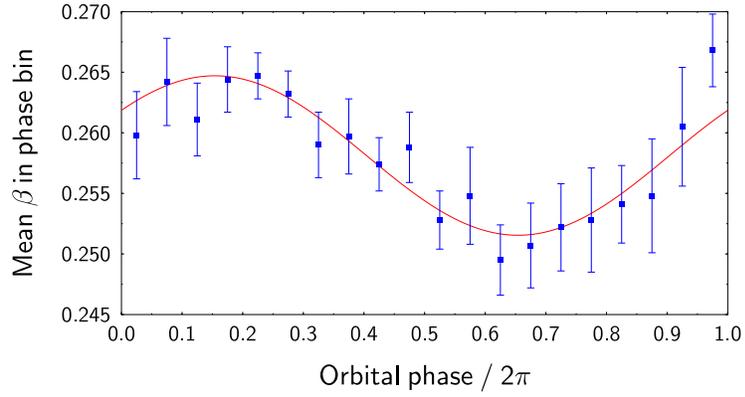}
\figcaption{\label{fig:fold} Speed data (Eqn\,\ref{eq:sum}) folded
over the orbital period of 13.08 days showing a clear sinusoid, mean
$0.2581 \pm 0.0005$, amplitude $0.0066 \pm 0.0007$, phase offset with
respect to optical ephemeris $2.17 \pm 0.11$\,rad.  }
\end{figure}

\begin{figure}
\epsscale{0.7}
\plotone{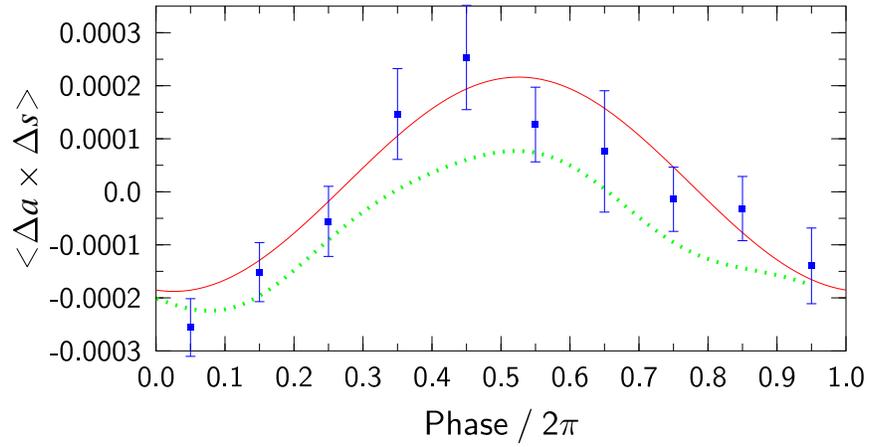}
\figcaption{\label{fig:a_s} The mean value of $\langle \Delta a \times
  \Delta s \rangle$ in 162-day period phase bins, showing the best fit
  to Eqn\,\ref{eq:as} averaged.  There are two free parameters in the
  fit shown as a solid line, $\langle \Delta\beta\,\Delta\theta
  \rangle$ and $\langle \Delta\beta\,\Delta\phi \rangle$. The dotted
  line shows the fit with fixed velocity but angular asymmetry.  }
\end{figure}

\begin{figure}
\epsscale{1.0}
\plotone{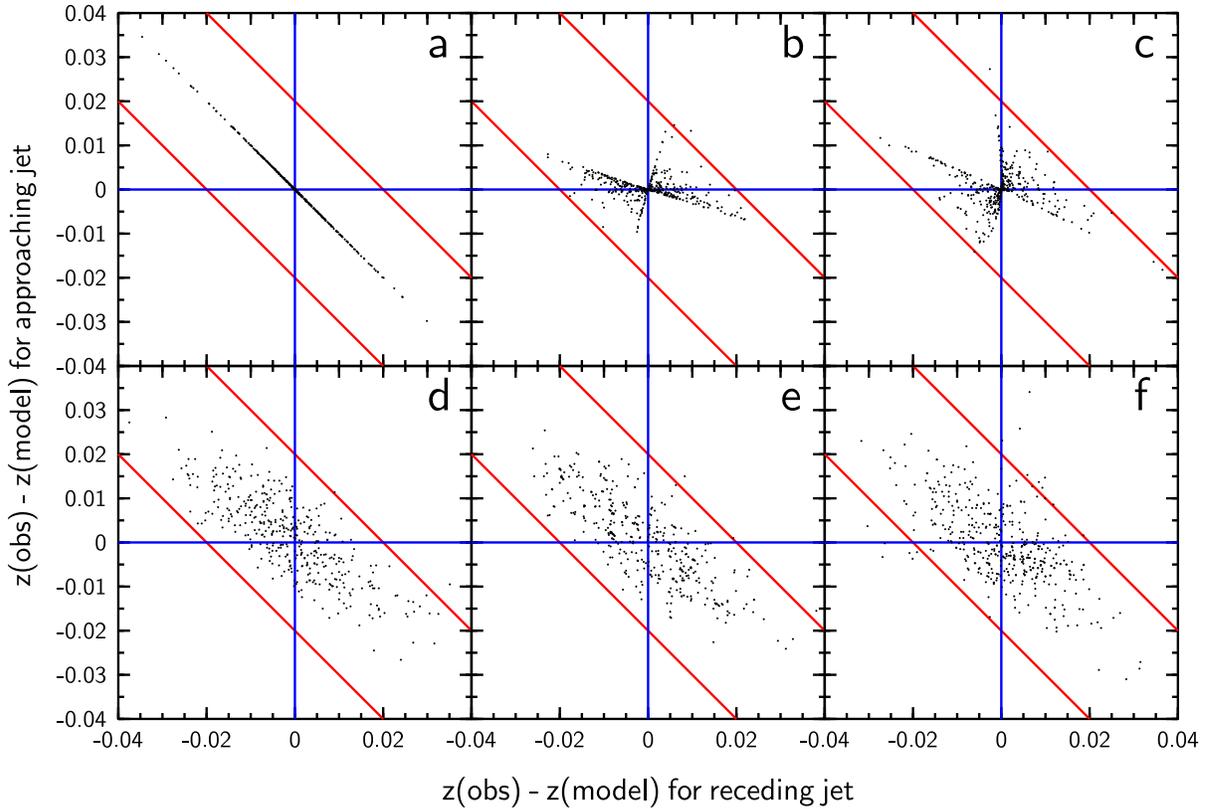}
\figcaption{\label{fig:eikfigfive} The effect on redshift residuals in
  each jet for deviations from the kinematic model in $\theta$,
  $\beta$ and $\phi$, for each date on which there is a redshift pair
  (replicating the sampling function).  {\bf (a)} Variations only in
  $\theta$, drawn from a gaussian distribution with half-width 2.7
  degrees.  The form of this plot, points lying only on the $y = -x$
  line, is identical for variations only in $\phi$, or for variations
  in both $\phi$ and $\theta$.  {\bf (b)} Speed-only variations (on a
  given day $\beta$ is drawn from a gaussian of half-width 0.013 ---
  Fig\,\ref{fig:resids_v_time}c). {\bf (c)} Variations in $\beta$ {\em
  anti-correlated} with those in $\theta$: on a given synthesized
  observation date the same randomly-drawn number from a gaussian is
  scaled by 0.013 for the speed variation and by $-2.7$ degrees for
  the $\theta$ variation.  {\bf (d)} As (c), but with $\theta$
  partially anti-correlated with $\beta$ and uncorrelated $\phi$
  fluctuations (as Table\,\ref{tab:fits}). {\bf (e)} As (d), but the
  duration of the $\beta$-variations, and that of the correlated
  $\theta$-component, is drawn from a gaussian of half-width
  2\,days. {\bf (f)} Observed residuals. }
\end{figure}

Archival optical spectroscopic data on SS\,433 reveal variations in
jet speed, in cone opening angle, and in the phase of the precession.
These appear in the plane of redshift residuals
(Fig\,\ref{fig:eikfigfive}) and through the new (symmetry dependent)
technique of combining simultaneously observed redshift pairs for the
speed-only ($s$) and angular-only ($a$) characteristics
(Fig\,\ref{fig:resids_v_time}).  The velocity variations $\sim
0.014\,c$ are strongly anticorrelated with cone angle $\theta$, in the
sense that when faster bolides are ejected the cone angle is smaller.
We also found smaller amplitude sinusoidal oscillations in speed as a
function of orbital phase.  If this is due to ejection speed, perhaps
the orbit of the binary is eccentric.  If these 13.08-day oscillations
are orbital Doppler shifts the orbital velocity is $\sim 400\,{\rm
km\, s}^{-1}$ (twice that inferred by \cite{Cra81,Fab90}). If this
were the case, the mass of the companion to SS\,433 would be $> 86\,
{\rm M}_\odot$ and if the mass fraction were 0.1, then the mass of the
companion would be \gtsim\ $100\, {\rm M}_\odot$.  Such masses would
be hardly consistent with an A-type companion \citep{Gei02,Cha04}.

\acknowledgments

K.M.B.\ thanks the Royal Society for a University Research Fellowship.
  It is a pleasure to thank
Avinash Deshpande, James Binney \& Philipp Podsiadlowski for helpful
discussions.

\clearpage

\begin{deluxetable}{llll}
\tablecaption{\label{tab:fits} 
Fits to excursions from the standard kinematic model (incorporating
nodding).
$\beta$ is in units of $c$, and $\theta$ and $\phi$ are in radians.}
\tablehead{\colhead{Quantity} & \colhead{fitted value} 
& \colhead{quantity} & \colhead{derived value}  \\ }
\startdata
$\langle \Delta \beta^2 \rangle     $  
& $\phantom{-}1.67 \pm 0.18 \times 10^{-4}$
& rms speed variation 
& 0.0129$c$
\\
$\langle \Delta \theta^2 \rangle    $  
& $\phantom{-}2.24 \pm 0.35 \times 10^{-3}$
& rms $\theta$ variation
& 2.71\,deg
\\
$\langle \Delta \phi^2 \rangle      $  
& $\phantom{-}1.31 \pm 0.26 \times 10^{-2}$
& rms $\phi$ variation
& 2.96\,days
\\
$\langle \Delta \beta\Delta\theta \rangle $  
&$-3.81 \pm 0.52 \times 10^{-4}$ &  & \\
$\langle \Delta \beta\Delta\phi \rangle   $  
& $\phantom{-}1.70 \pm 0.80 \times 10^{-4}$ &  & \\
$\langle \Delta \theta\Delta\phi \rangle$
& indistinguishable from zero & & \\
\enddata
\end{deluxetable}

\clearpage

\clearpage

\clearpage

\clearpage

\clearpage

\clearpage

\end{document}